\documentclass{article}

\usepackage{arxiv}

\usepackage[utf8]{inputenc} 
\usepackage[T1]{fontenc}    
\usepackage{hyperref}       
\usepackage{url}            
\usepackage{booktabs}       
\usepackage{amsfonts}       
\usepackage{nicefrac}       
\usepackage{microtype}      
\usepackage{lipsum}		
\usepackage{graphicx}
\usepackage{natbib}
\usepackage{doi}
\usepackage{amsmath} 
\usepackage{amssymb}
\usepackage{xcolor}

\title{Theoretical prediction of Structural Stability and Superconductivity in Janus Ti$_2$CSH MXene}

\author{ \href{https://orcid.org/0009-0004-2196-8245}{\includegraphics[scale=0.06]{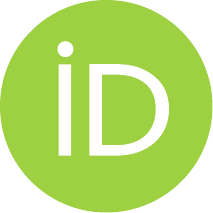}\hspace{1mm}Jakkapat Seeyangnok$^{*}$} \\
	Department of Physics\\
    Faculty of Science\\
	Chulalongkorn University\\
	Bangkok, Thailand \\
	\texttt{jakkapatjtp@gmail.com} \\
	\And
	\href{https://orcid.org/0000-0002-8450-7751}{\includegraphics[scale=0.06]{orcid.pdf}\hspace{1mm}Udomsilp Pinsook} \\
	Department of Physics\\
    Faculty of Science\\
	Chulalongkorn University\\
	Bangkok, Thailand \\
	\texttt{Udomsilp.P@Chula.ac.th} \\
}

\begin{document}
\maketitle

\begin{abstract}
    We present a comprehensive first-principles investigation of the structural stability, vibrational characteristics, and superconducting properties of the Janus Ti$_2$CSH monolayer.  Janus MXene (JMXene) materials, such as Ti$_2$CSH, have attracted significant attention due to their intrinsic two-dimensional structure and the absence of out-of-plane symmetry, which give rise to novel physical phenomena. Phonon calculations confirm the dynamical stability of the monolayer, while electronic structure and electron-phonon coupling analyses reveal a strong phonon-mediated pairing mechanism. Anisotropic Migdal-Eliashberg theory predicts a single-gap superconducting state, with gap values between 4.29 and 4.71~meV at 10~K and a critical temperature ($T_c$) of 22.6~K. These findings establish Ti$_2$CSH as a promising two-dimensional superconductor with potential applications in quantum and nanoscale technologies.
\end{abstract}

\keywords{Superconductivity \and 2D materials \and Janus MXene \and Electron-phonon coupling.}

	\section{Introduction}\label{sec1}
    Since graphene’s isolation in 2004~\cite{novoselov2004electric}, two-dimensional (2D) materials have emerged as a dynamic frontier in condensed matter physics and nanotechnology, owing to their novel quantum phenomena and wide-ranging technological prospects. At the same time, the discovery of superconductivity~\cite{frohlich1950theory,migdal1958interaction,eliashberg1960interactions,nambu1960quasi,allen1975transition,pinsook2024analytic} in high-pressure hydrides has revolutionized the search for materials with high critical temperatures, largely driven by strong electron-phonon interactions, as originally suggested for metallic hydrogen and its compounds by Ashcroft~\cite{ashcroft1968metallic, ashcroft2004hydrogen}. These concurrent advances have inspired efforts to bridge both material classes—2D systems and superconducting hydrides—in pursuit of superconductivity at reduced or ambient pressures. Among the most promising candidates are 2D hydrogen-rich materials, which offer fertile ground for conventional phonon-mediated superconductivity. Theoretical studies initially proposed that fully hydrogenated graphene, known as graphane, could display superconductivity with critical temperatures ($T_c$) above 90~K~\cite{sofo2007graphane}. Following this, several ternary 2D hydrides were introduced, including hydrogen-functionalized MgB$_2$ monolayers with predicted $T_c$ near 67~K~\cite{savini2010first}, and hydrogenated HPC$_3$ with $T_c$ around 31~K~\cite{li2022phonon} along with other works of hydrogenated 2D materials ~\cite{bekaert2019hydrogen,han2023high,liu2024three,seeyangnok2025high_npj2d,seeyangnok2025hydrogenation_nanoscale}

    Transition metal dichalcogenides (TMDs) represent another crucial family of 2D compounds characterized by their van der Waals stacking and rich phase diversity. These materials are known for supporting electronic flat bands~\cite{calandra2018phonon, rostami2018helical, tresca2019charge, nakata2021robust}, which can be further engineered via heterostructuring or twist-angle manipulation~\cite{zhang2020flat, vitale2021flat, kuang2022flat, rademaker2022spin, huang2023recent}. A notable subset of TMDs are Janus monolayers (JMXenes), which exhibit broken out-of-plane symmetry due to the substitution of one chalcogen layer with a different atomic species. This structural asymmetry introduces new degrees of freedom for tailoring mechanical, optical, and electronic behavior~\cite{tang20222d, varjovi2021janus, zhang2022janus, angeli2022twistronics, maghirang2019predicting, he2018two, yeh2020computational, yin2021recent, li2023structure}. Although absent in nature, Janus structures have been synthesized successfully, beginning with Janus graphene in 2013~\cite{zhang2013janus}. Since then, a range of Janus TMDs such as MoSSe, WSSe, and PtSSe have been experimentally realized~\cite{trivedi2020room, lu2017janus, sant2020synthesis}, primarily through selective atomic substitution methods like SEAR (Selective Epitaxy Atomic Replacement) using hydrogen plasma~\cite{trivedi2020room, tang20222d}. 

    Hydrogenation continues to play a pivotal role in modifying the electronic and magnetic characteristics of 2D transition metal dichalcogenides (TMDs). Recently, a Janus 2H-MoSH monolayer (ML) was synthesized using the SEAR method by substituting the top sulfur atoms with hydrogen atoms~\cite{lu2017janus}. Theoretically, the Janus MoSH ML has been predicted to exhibit superconductivity with a critical temperature ($T_c$) of 26.81~K in the 2H phase~\cite{liu2022two,ku2023ab} along with its bilayer properties ~\cite{pinsook2025superconductivity}. This has inspired subsequent studies, including MoSLi~\cite{xie2024strong} and MoSeLi~\cite{moseliseeyangnok}, where two-gap superconductivity has been predicted. Furthermore, the 2H and 1T phases of WSH and WSeH have been suggested to be dynamically stable with $T_c$ values exceeding 12~K~\cite{wseh_prb,wsh_2dmat}. This phase stability and superconducting behavior have been corroborated by subsequent independent studies~\cite{gan2024hydrogenation,fu2024superconductivity}. Moreover, there have been numerous investigations into the physical properties of group IV transition metal dichalcogenides (MX$_2$, M = Ti, Zr, Hf; X = S, Se, Te). For an incomplete list, see Refs.~\cite{joseph2023review,lasek2021synthesis,mattinen2019atomic,zhang2016systematic,toh2016catalytic,xie2015two}. We expect that the SEAR method will also enable the synthesis of Janus transition metal chalcogenide (JTMC) hydrides, for which superconductivity has been predicted in several phases: 2H-TiSH, 2H-TiSeH, 1T-TiSH, 1T-TiSeH, 1T-TiTeH, and 1T-ZrTeH, with $T_c$ values of 13.09~K, 30.19~K, 18.30~K, 13.31~K, 9.04~K, respectively, in the 10–30~K range~\cite{li2024machine,ul2024superconductivity}. However, investigations also show that Janus MXH (M = Ti, Zr, Hf; X = S, Se, Te) structures exhibit magnetic ordering as the ground state ~\cite{seeyangnok2025competition}, similar to CrSH~\cite{sukserm2025half}, highlighting their multifunctional potential~\cite{yan2022enhanced,han2023theoretical,xue2024realization}.

    In this work, we present a comprehensive first-principles investigation of a novel class of Janus transition metal hydride monolayers, Ti$_2$CSH, formed by hydrogen substitution of sulfur in Ti$_2$CS$_2$, which has been extensively studied~\cite{yuan2021multilayer,zhang2022first,wang2023theoretical}. Motivated by the successful synthesis of a Janus 2H-MoSH monolayer via the SEAR method—achieved by substituting the top sulfur atoms in MoS$_2$ with hydrogen~\cite{lu2017janus}—we explore the thermodynamic stability, electronic structure, vibrational modes, and superconducting characteristics of Ti$_2$CSH. Our findings provide new insights into the design of two-dimensional superconductors and highlight the potential of Janus hydride materials for achieving high-temperature superconductivity under experimentally feasible conditions.

	\section{Computational details}
We employed density functional theory (DFT) within the \textsc{Quantum ESPRESSO} (QE) package~\cite{giannozzi2009quantum,giannozzi2017advanced} to investigate the material’s properties. The crystal structure was constructed in \textsc{VESTA}~\cite{momma2011vesta} with trigonal symmetry \( P\overline{3}m1 \) (No.~156).  

Computational parameters were carefully chosen to balance accuracy with efficiency. A plane-wave cutoff of 80~Ry and a charge-density cutoff of 320~Ry were applied. The Brillouin zone was sampled using a $24 \times 24 \times 1$ Monkhorst–Pack mesh~\cite{monkhorst1976special}. Electronic occupations were treated with a Methfessel–Paxton smearing of 0.02~Ry~\cite{methfessel1989high}. For the ionic potential, we used optimized norm-conserving Vanderbilt pseudopotentials~\cite{hamann2013optimized,schlipf2015optimization} together with the generalized gradient approximation (GGA) in the Perdew–Burke–Ernzerhof (PBE) formulation~\cite{perdew1996generalized}.  

The atomic structure was relaxed using the Broyden–Fletcher–Goldfarb–Shanno (BFGS) minimization algorithm~\cite{BFGS,liu1989limited} until residual forces were below \(10^{-5}~\text{eV/\AA}\). Phonon properties were computed via density functional perturbation theory (DFPT) on a $12 \times 12 \times 1$ grid. Electron–phonon coupling gives rise to finite phonon lifetimes, expressed as a linewidth \(\gamma_{\boldsymbol{q}\nu}\):  

\begin{equation}
\gamma_{\boldsymbol{q}\nu} = 2\pi\omega_{\boldsymbol{q}\nu}\sum_{nm}\sum_{\boldsymbol{k}} \left| g_{\boldsymbol{k}+\boldsymbol{q},\boldsymbol{k}}^{\boldsymbol{q}\nu,mn} \right|^2 \delta(\epsilon_{\boldsymbol{k}+\boldsymbol{q},m}-\epsilon_F) \delta(\epsilon_{\boldsymbol{k},n}-\epsilon_F),
\label{gammaphononlinewidths}
\end{equation}

where \( g_{\boldsymbol{k}+\boldsymbol{q},\boldsymbol{k}}^{\boldsymbol{q}\nu,mn} \) is the electron–phonon matrix element and \(\omega_{\boldsymbol{q}\nu}\) is the phonon frequency. The corresponding electron–phonon coupling strength for a given mode is:  

\begin{equation}
\lambda_{\boldsymbol{q}\nu} = \frac{\gamma_{\boldsymbol{q}\nu}}{\pi N(\epsilon_F) \omega_{\boldsymbol{q}\nu}^2},
\label{eqn:lambda_qv}
\end{equation}

with \( N(\epsilon_F) \) denoting the electronic density of states at the Fermi level.  

To examine superconductivity, we solved the Migdal–Eliashberg equations \cite{bardeen1957microscopic,frohlich1950theory,migdal1958interaction,eliashberg1960interactions,nambu1960quasi,pinsook2024analytic} using the EPW code~\cite{noffsinger2010epw,ponce2016epw}, based on the Wannier–Fourier interpolation scheme of Giustino \textit{et al.}~\cite{giustino2007electron,giustino2017electron}. EPW self-consistently determines the superconducting gap function \(\Delta_{nk}(i\omega_j)\) and renormalization function \(Z_{nk}(i\omega_j)\) on the Matsubara axis, with frequencies \(\omega_j = (2j+1)\pi T\). The Coulomb pseudopotential was fixed at \(\mu^* = 0.1\). For convergence, dense k- and q-point meshes of \(160 \times 160 \times 1\) and \(80 \times 80 \times 1\) were adopted. A Fermi surface broadening of 0.48~eV and a Matsubara cutoff of 1.20~eV were applied. Gaussian smearing was used for delta functions, with widths of 0.12~eV for electrons and 0.5~meV for phonons.  

The Migdal–Eliashberg equations read:  

\begin{equation}
Z_{nk}(i\omega_j) = 1 + \frac{\pi T}{N(\varepsilon_F)\omega_j} \sum_{mk'j'} \frac{\omega_{j'}}{\sqrt{\omega_{j'}^2 + \Delta_{mk'}^2(i\omega_{j'})}},
\label{eqn-ME1}
\end{equation}

\begin{eqnarray}
Z_{nk}(i\omega_j)\Delta_{nk}(i\omega_j) &= \frac{\pi T}{N(\varepsilon_F)} \sum_{mk'j'} \frac{\Delta_{mk'}(i\omega_{j'})}{\sqrt{\omega_{j'}^2 + \Delta_{mk'}^2(i\omega_{j'})}} \delta(\epsilon_{mk'} - \varepsilon_F) \nonumber \\
&\times \left[\lambda(nk, mk', \omega_j - \omega_{j'}) - \mu^*\right].
\label{eqn-ME2}
\end{eqnarray}

Finally, to assess the mechanical stability of the Janus Ti$_2$CSH monolayer, we calculated its in-plane elastic constants, characteristic of 2D hexagonal materials. These constants \(C_{ij}\) were obtained from the second derivative of the total energy with respect to strain:  

\begin{equation}
C_{ij} = \frac{1}{S_0} \frac{\partial^2 E}{\partial \epsilon_i \partial \epsilon_j},
\end{equation}

where \(S_0\) is the equilibrium unit-cell area, and \(\epsilon_i\), \(\epsilon_j\) are the applied strain components.  

In 2D hexagonal systems, only two constants are independent, \(C_{11}\) and \(C_{12}\), with the shear modulus defined as \(C_{66} = \tfrac{1}{2}(C_{11} - C_{12})\). The elastic response under small deformations is then described by the two-dimensional Hooke’s law:  

\begin{equation}
\sigma =
\begin{bmatrix}
C_{11} & C_{12} & 0 \\
C_{12} & C_{22} & 0 \\
0 & 0 & C_{66}
\end{bmatrix}
\varepsilon.
\end{equation}  

\section{Structural properties}
	\subsection{Crystal structure}
	\begin{figure}[h]
		\centering
		\includegraphics[width=16cm]{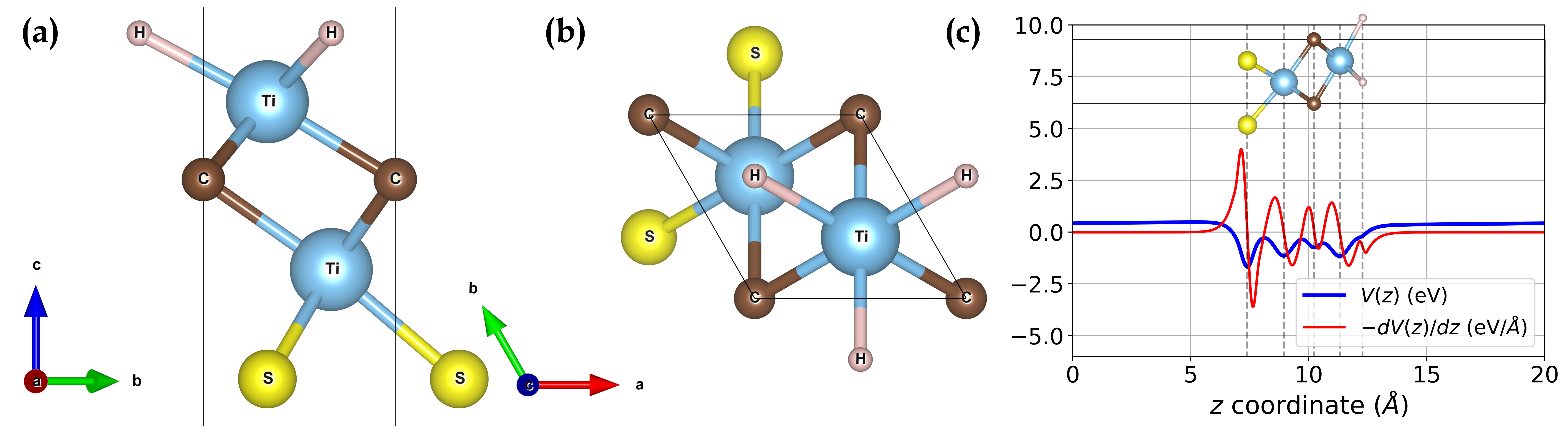}
		\caption{(a) Side view and (b) top view of the optimized atomic structure of the Janus Ti$_2$CSH monolayer, illustrating the asymmetric stacking of atoms within the trigonal $P\overline{3}m1$ space group. Titanium (Ti), carbon (C), sulfur (S), and hydrogen (H) atoms are represented by blue, brown, yellow, and pink spheres, respectively. The lattice vectors $\vec{a}$, $\vec{b}$, and $\vec{c}$ are indicated for clarity. (c) Planar-averaged electrostatic potential $V(z)$ (blue) and corresponding electric field $-dV(z)/dz$ (red) along the out-of-plane direction $z$. The asymmetry of the potential profile confirms the existence of an intrinsic dipole moment across the monolayer, although the nearly flat region in the vacuum implies a negligible net electric field beyond the layer.}
		\label{fig:structure}
	\end{figure}

The Janus Ti$_2$CSH monolayer adopts a periodically repeated hexagonal supercell that crystallizes in the trigonal space group \( P\overline{3}m1 \) (No. 156), as illustrated in Figure~\ref{fig:structure}. In this structure, the X atom—either carbon or nitrogen—is located at the Wyckoff position \((0,0)\), while the sulfur atoms occupy the \((1/3, 2/3)\) and \((2/3, 1/3)\) sites in the $xy$-plane, representing the lower and upper sublayers, respectively. The in-plane lattice constant is denoted by $a$. Hydrogen atoms can, in principle, be positioned at either \((1/3, 2/3)\) or \((2/3, 1/3)\), corresponding to 2H- and 1T-like stacking configurations, respectively—similar to those observed in transition metal dichalcogenides (TMDs). Our total energy calculations reveal that the 1T phase is energetically more favorable than the 2H phase, with energy differences of 0.43~eV. We also analyzed the possible magnetic ground states of Ti$_2$CSH, including ferromagnetic (FM) and antiferromagnetic configurations such as G-type (GAF) and C-type (CAF) ordering. Among these possibilities, the non-magnetic (NM) metallic phase is found to be the most stable.

To assess the emergence of an intrinsic electric field resulting from the inherent asymmetry in Ti$_2$CSH, we examined the planar-averaged electrostatic potential across the monolayer. As depicted in Figure~\ref{fig:structure}~(c), the structural non-centrosymmetry produces a noticeable potential offset between the upper and lower atomic layers. This internal potential variation, stemming from the geometric asymmetry, leads to the formation of an intrinsic out-of-plane electric field. While a slight potential gradient is observed across the vacuum region, the near-flat profile of the electrostatic potential within the monolayer indicates that any net internal electric field is negligibly small. This validates the reliability of our DFT-based simulations for the Ti$_2$CSH monolayer system.
    \subsection{Energetic stability}

To evaluate the thermodynamic viability of synthesizing the Janus Ti$_2$CSH monolayer, we computed its formation energy ($E_{\text{formation}}$) using elemental bulk phases as reference points~\cite{zhang1991chemical,kirklin2015open}. The formation energy is defined as:

\begin{equation}
    E_{\text{formation}} = \frac{5E_{\text{Ti$_2$CSH, 2D}} - 2E_{\text{Ti, bulk}} - E_{\text{C, bulk}} - E_{\text{S, bulk}} - E_{\text{H$_2$, bulk}}}{5},
\end{equation}

where $E_{\text{Ti$_2$CSH, 2D}}$ denotes the energy per atom of the relaxed Ti$_2$CSH monolayer, and $E_{\text{Ti, bulk}}, E_{\text{C, bulk}}, E_{\text{S, bulk}},$ and $E_{\text{H$_2$, bulk}}$ refer to the energies per atom of bulk Ti (in the hexagonal P6$_3$/mmc structure), diamond-structured C, orthorhombic $\alpha$-S$_8$ (space group \textit{Fddd})~\cite{fedyaeva2023stability}, and molecular hydrogen, respectively. The calculated formation energy is \(-0.05826~\text{eV}\) or \(-5.619~\text{kJ/mol}\) per atom, which lies within the typical range for bulk metal diborides~\cite{colinet2018enthalpies}.

The substantially negative values of both the cohesive and formation energies indicate that the Janus Ti$_2$CSH monolayer is thermodynamically stable. These results suggest that experimental synthesis may be feasible using conventional techniques such as chemical vapor deposition (CVD).

    \subsection{Thermal stability}  
    \begin{figure}[h!]
		\centering
		\includegraphics[width=16cm]{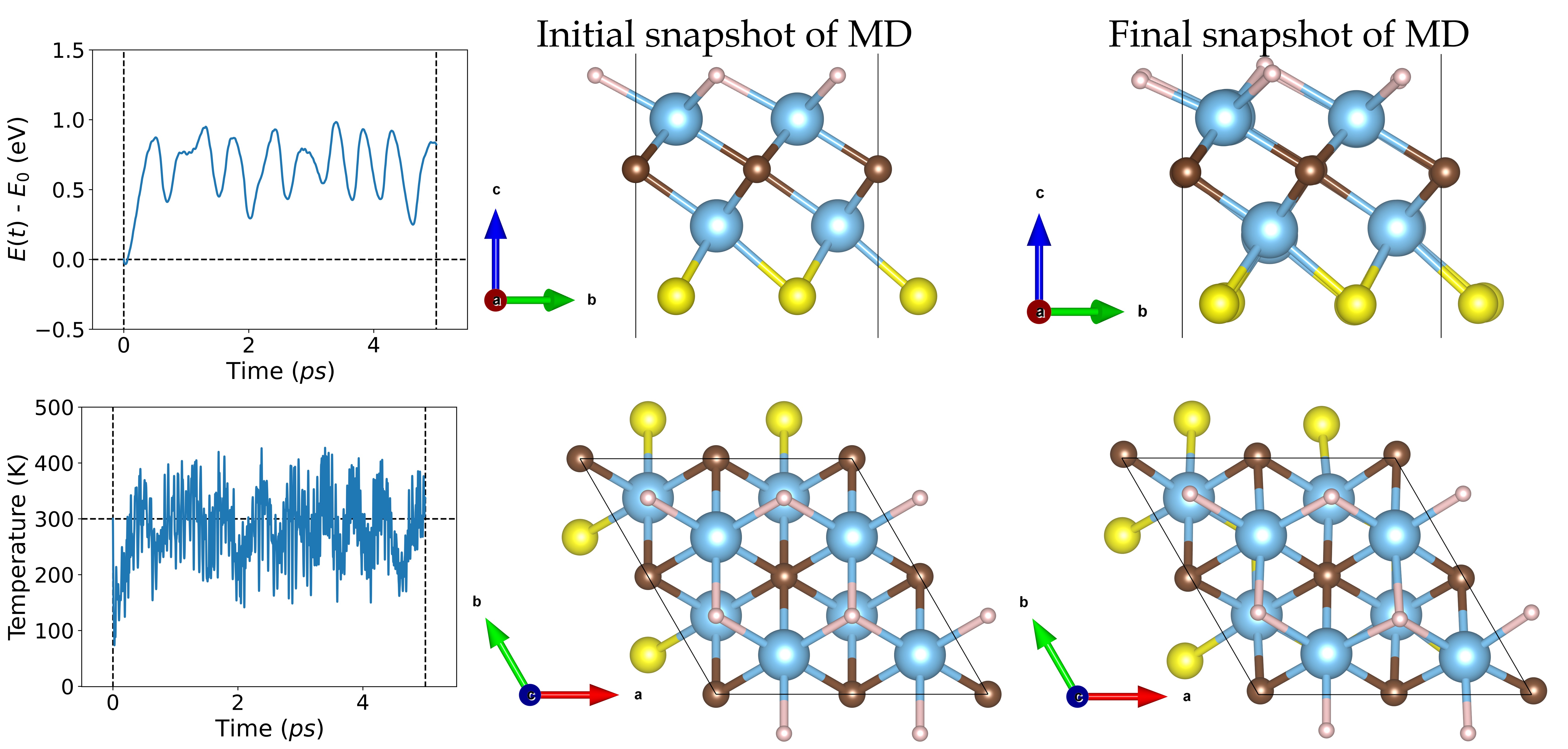}
		\caption{Thermal stability of the Janus Ti$_2$CSH monolayer assessed via \textit{ab initio} molecular dynamics (AIMD) simulations at room temperature over a 5~ps timescale. (a) Time evolution of the total energy relative to the initial value, showing energy fluctuations indicative of thermal equilibrium. (b) Time evolution of the system temperature, fluctuating around the target value of 300~K, confirming thermalization. (c) and (d) Top and side views of the atomic structure at the beginning and end of the simulation, respectively. The structural configuration remains essentially unchanged, indicating that Ti$_2$CSH retains its integrity under thermal perturbation.}
  	\label{fig:MD-plos}
	\end{figure}
    To evaluate the thermal robustness of the Janus Ti$_2$CSH monolayer beyond the limitations of harmonic approximations inherent in phonon and elastic analyses, we carried out \textit{ab initio} molecular dynamics (AIMD) simulations. Unlike static methods, AIMD provides insight into the real-time response of the atomic lattice at finite temperatures, allowing us to probe potential thermal distortions, phase transitions, or surface reconstructions that might arise under ambient conditions.

    The simulations were conducted using a $2 \times 2 \times 1$ supercell containing 20 atoms within the canonical (NVT) ensemble framework. The system was equilibrated at room temperature and propagated for 10,000 steps with a time increment of 0.5~fs per step. As shown in Figure~\ref{fig:MD-plos}, the atomic configuration remained structurally stable, with no signs of degradation or reconstruction, and the total energy exhibited minor fluctuations around a steady-state value. These findings confirm that Ti$_2$CSH retains its structural integrity under thermal perturbations, supporting the dynamical stability inferred from phonon calculations.

    \subsection{Mechanical stability} 
    Our first-principles calculations reveal that the Janus Ti$_2$CSH monolayer possesses in-plane stiffness values of \(C_{11} = C_{22} = 12.83~\text{eV/\AA}^2\) and \(C_{12} = 5.62~\text{eV/\AA}^2\), indicating isotropic elastic properties. These values satisfy the Born mechanical stability criteria for 2D materials: \(C_{11}C_{22} - C_{12}^2 > 0\), along with the positivity of \(C_{11}\), \(C_{22}\), and \(C_{66}\), as discussed in the stability framework by Mouhat and Coudert~\cite{mouhat2014necessary}. This confirms that the Ti$_2$CSH monolayer is mechanically stable.

    Taken together with the absence of imaginary modes in the phonon spectrum, the thermal resilience demonstrated by AIMD simulations, and the favorable energetic properties such as negative formation and substantial cohesive energies, our findings support the potential of Ti$_2$CSH as a viable and robust monolayer material for experimental realization and technological applications.


	
    	\section{Electronic properties}
   \begin{figure}[h!]
		\centering
		\includegraphics[width=11cm]{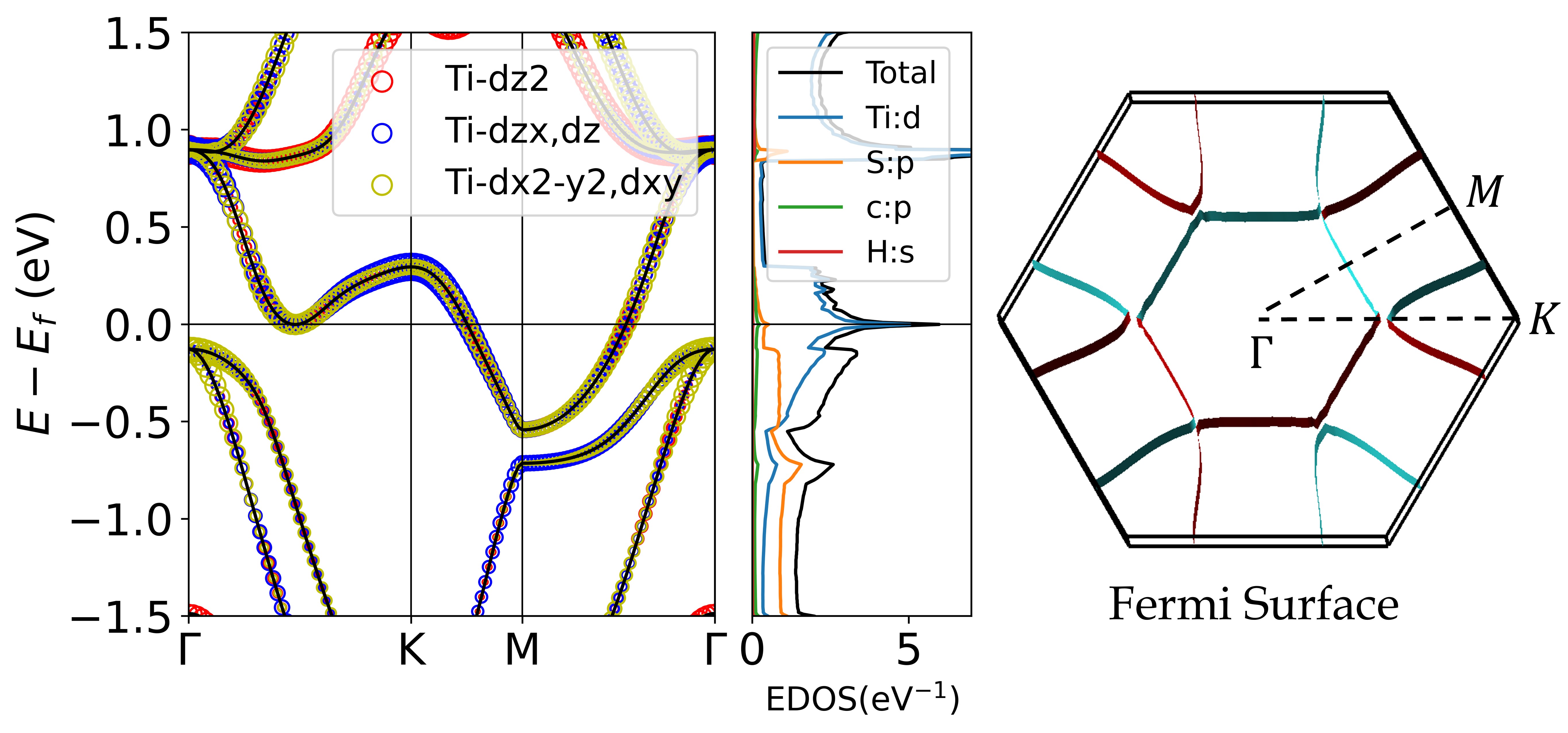}
		\caption{Left:Electronic band structure and projected density of states (PDOS) of the Ti$2$CSH monolayer. The left panel shows the orbital-resolved band structure with Ti-$d$ orbital contributions highlighted: $d{z^2}$ (red), $d_{xz}$/$d_{yz}$ (blue), and $d_{x^2-y^2}$/$d_{xy}$ (yellow). The right panel displays the total and element-projected DOS, indicating the dominant role of Ti-$d$ orbitals near the Fermi level ($E_F = 0$ eV). The electronic structure confirms the metallic character of the material, with multiple band crossings at the Fermi energy. Right: Visualization of the Fermi surface formed by electronic bands intersecting the Fermi level.}
		\label{fig:electronics}
    \end{figure}    	
    Figure~\ref{fig:electronics} displays the electronic properties of the Ti$_2$CSH monolayer, including the orbital-projected band structure, total and partial density of states, and the Fermi surface within the Brillouin zone. The results indicate that Ti$_2$CSH exhibits metallic characteristics, as evidenced by the presence of bands crossing the Fermi energy. These bands are predominantly contributed by Ti-\(d\) orbitals.

    Notably, a band crossing occurs along the $\Gamma$--$M$ direction, giving rise to a Fermi contour situated between these two high-symmetry points. In addition, further crossings appear along the $\Gamma$--$K$ path, resulting in two distinct Fermi surfaces: one encircling the $\Gamma$ point and another centered near the $M$ point. Both features are primarily composed of Ti-\(d\) orbital character, as shown in Figure~\ref{fig:electronics}.

    The topology of the Fermi surface, shaped by these crossings, plays a crucial role in the formation of superconducting Cooper pairs. As demonstrated in Figure~\ref{fig:epw-gap}, the superconducting gap inherits this Fermi surface topology, indicating that the pairing mechanism is closely linked to the underlying electronic structure at the Fermi level through phonon momentum exchange.
	\section{Phonons}
    \begin{figure}[h!]
		\centering
		\includegraphics[width=16cm]{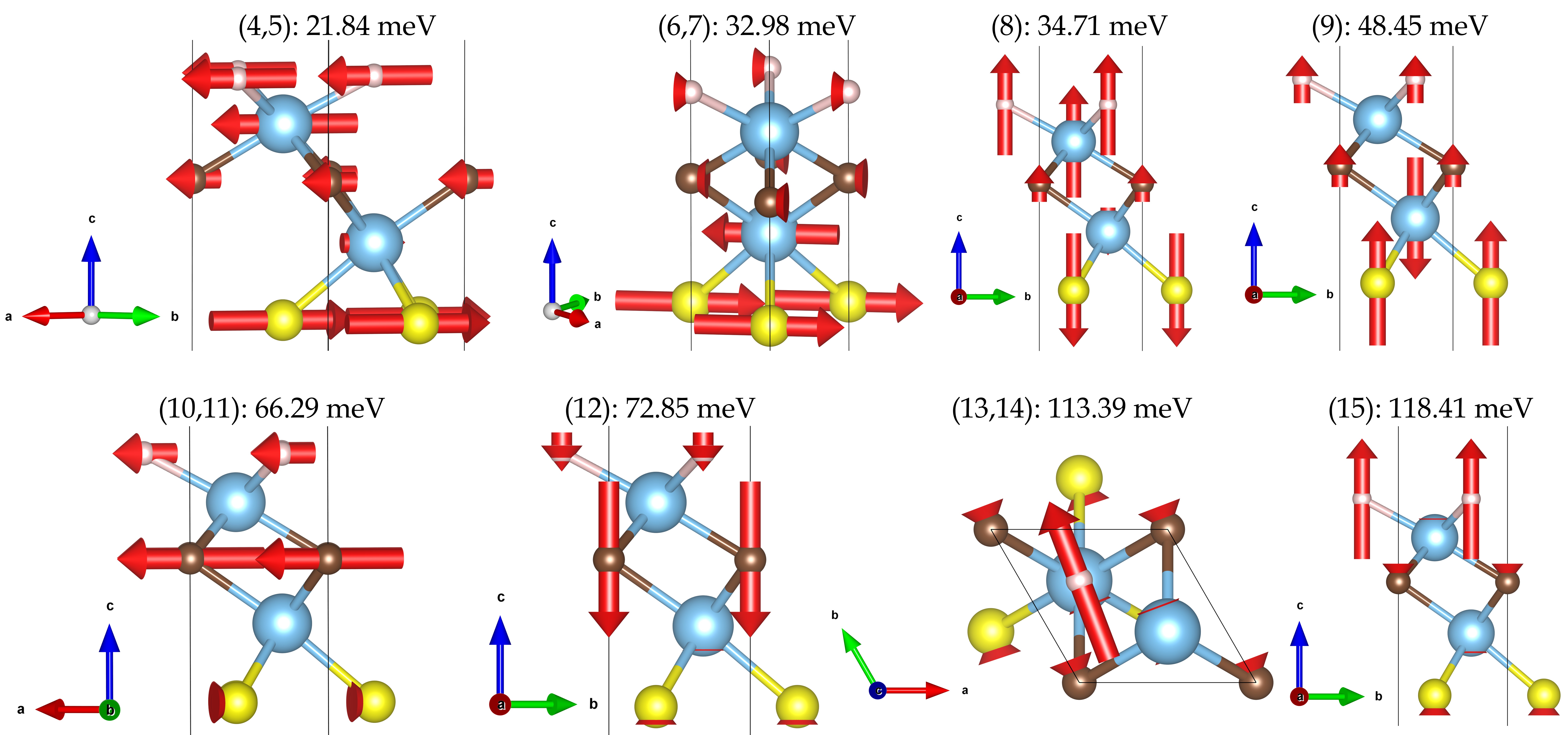}
		\caption{Figures show the visualization of the vibrational modes at the $\Gamma$ point corresponding to the optical phonon modes listed in Table~\ref{tab:phonon-eigenvalues}. Each mode illustrates the relative atomic displacements within the unit cell, providing insight into the character and symmetry of the phonon modes.}
		\label{fig:eigenvectors}
	\end{figure}

\begin{figure}[h!]
		\centering
		\includegraphics[width=10cm]{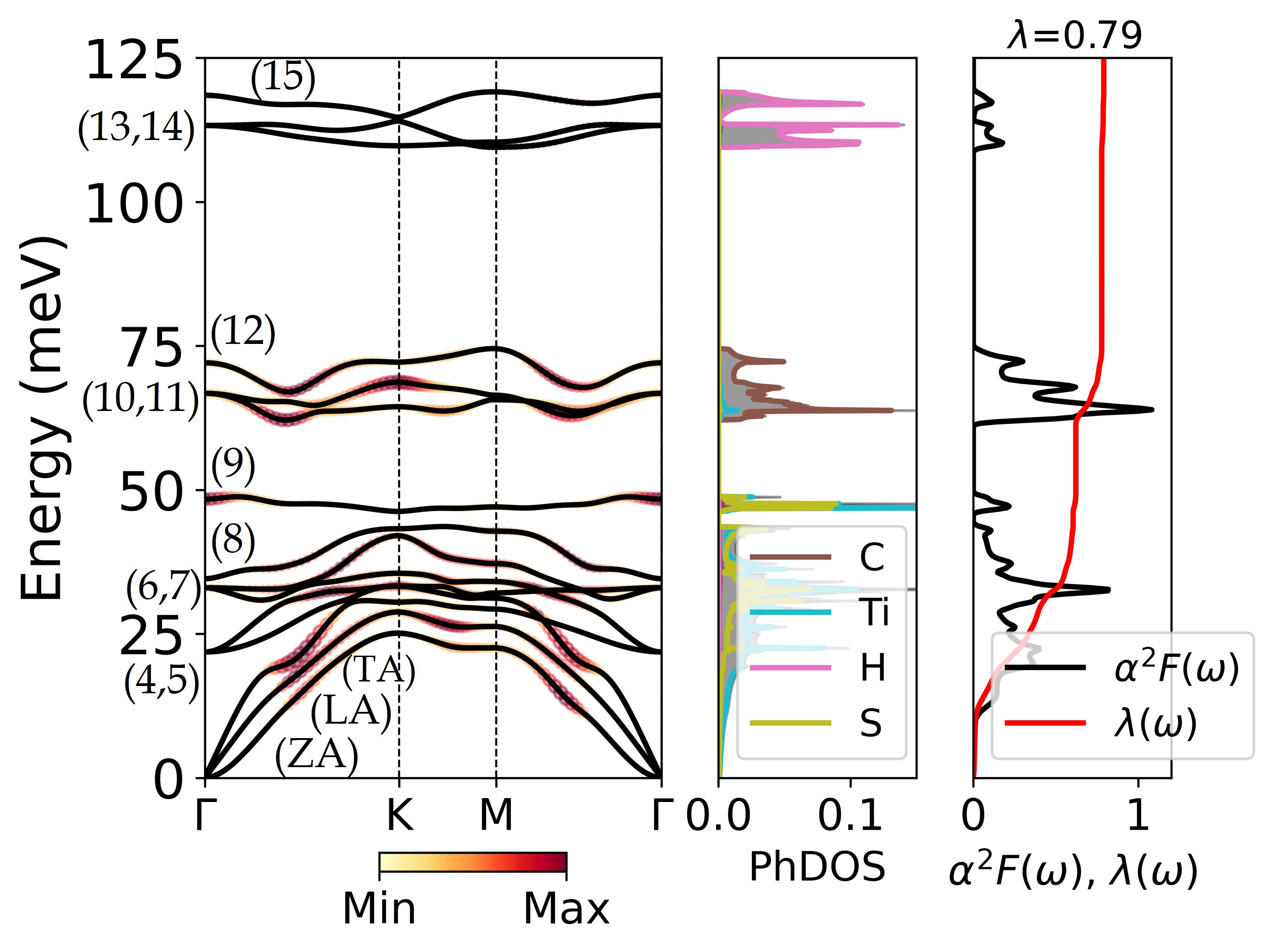}
		\caption{Phonon and electron-phonon interaction properties of Ti$_2$CSH monolayer. Left: Phonon dispersion curves overlaid with the electron-phonon coupling (EPC) strength, highlighted by the color gradient from yellow (low) to red (high). The phonon branches are labeled with their corresponding mode indices, and acoustic modes (ZA, LA, TA) are indicated near the $\Gamma$ point. Middle: Projected phonon density of states (PhDOS), showing atomic contributions from Ti, C, S, and H atoms. Right: Eliashberg spectral function $\alpha^2F(\omega)$ (black) and the integrated EPC constant $\lambda(\omega)$ (red), with a total EPC value of $\lambda = 1.04$, indicating strong electron-phonon coupling.}
		\label{fig:phonons}
	\end{figure}

    The phonon spectrum of Ti$_2$CSH, displayed in Figure~\ref{fig:phonons}, affirms the system’s dynamical stability through the absence of any imaginary frequencies across the Brillouin zone. At the $\Gamma$ point, the vibrational modes adhere to the $C_{3v}$ ($3m$) point group symmetry, comprising three acoustic and six optical branches. The acoustic modes consist of the longitudinal acoustic (LA) and transverse acoustic (TA) modes confined to in-plane motion, as well as the flexural acoustic (ZA) mode with out-of-plane character. Interestingly, the ZA mode reveals a softening near the $K$ point, identified by a characteristic dip, which implies enhanced electron-phonon interaction at that specific wavevector $\boldsymbol{q}$—a feature clearly captured in the electron-phonon coupling-weighted phonon dispersion overlay in Figure~\ref{fig:phonons}.

    The optical branches are categorized under the $E$ and $A_1$ irreducible representations, as outlined in Table~\ref{tab:phonon-eigenvalues}. Vibrational modes corresponding to bands (4,5), situated around 21.84~meV, arise from the collective in-plane oscillations of Ti, S, C, and H atoms. Modes (6,7) near 32.98~meV predominantly involve in-plane motion of S atoms. The 34.71~meV mode (8) is governed by out-of-plane vibrations of Ti, S, and H atoms, while mode (9) at 48.45~meV corresponds to out-of-plane displacements of Ti and S. The bands labeled (10,11) at 66.29~meV originate from planar vibrations of carbon atoms, and band (12) at 72.85~meV involves out-of-plane movement of carbon. High-frequency modes (13,14) at 113.39~meV are dominated by in-plane vibrations of hydrogen, and the uppermost band (15) at 118.41~meV corresponds to its out-of-plane motion. These mode patterns are visually represented in Figure~\ref{fig:eigenvectors}. When examining the phonon spectrum across the Brillouin zone, the vibrational modes can be broadly categorized into three distinct energy ranges based on atomic contributions. The low-energy region (0--50~meV) is primarily dominated by vibrations of Ti and S atoms. The intermediate range (60--75~meV) mainly arises from the vibrations of C atoms, while the high-energy region (110--120~meV) is predominantly associated with H atom vibrations. These trends are clearly illustrated in the atom-projected phonon density of states shown in Figure~\ref{fig:phonons}.

    Due to the lack of inversion symmetry in the monolayer’s lattice structure, all phonon modes exhibit both Raman and infrared activity. A detailed breakdown of the eigenfrequencies and mode characteristics is summarized in Table~\ref{tab:phonon-eigenvalues}.

        \begin{table}[h!]
    \centering
	\caption{\label{tab:phonon-eigenvalues}
Phonon mode characteristics at the $\Gamma$ point for the Ti$_2$CSH monolayer, including band indices, symmetry labels, infrared (IR) and Raman (R) activity, and the corresponding frequencies (in meV).}
	   \begin{tabular}{lclclclclcl}
        \hline
    Bands $\nu$ & Subgroup  & Active & energy (meV) \\
	   \hline
		4,5 & $E~L_3$ & I+R & 21.84 \\
		6,7 & $E~L_3$ & I+R & 32.98 \\
		8 & $A_1~L_1$ & I+R & 34.71 \\
		9 & $A_1~L_1$ & I+R & 48.45 \\
        10,11 & $E~L_3$ & I+R & 66.29 \\
        12 & $A_1~L_1$ & I+R & 72.85 \\
        13,14 & $E~L_3$  & I+R & 113.39 \\
        15 & $A_1~L_1$  & I+R & 118.41 \\
            \hline
	   \end{tabular}
	\end{table}

\section{Superconductivity}
    
    \begin{figure}[h!]
		\centering
		\includegraphics[width=12cm]{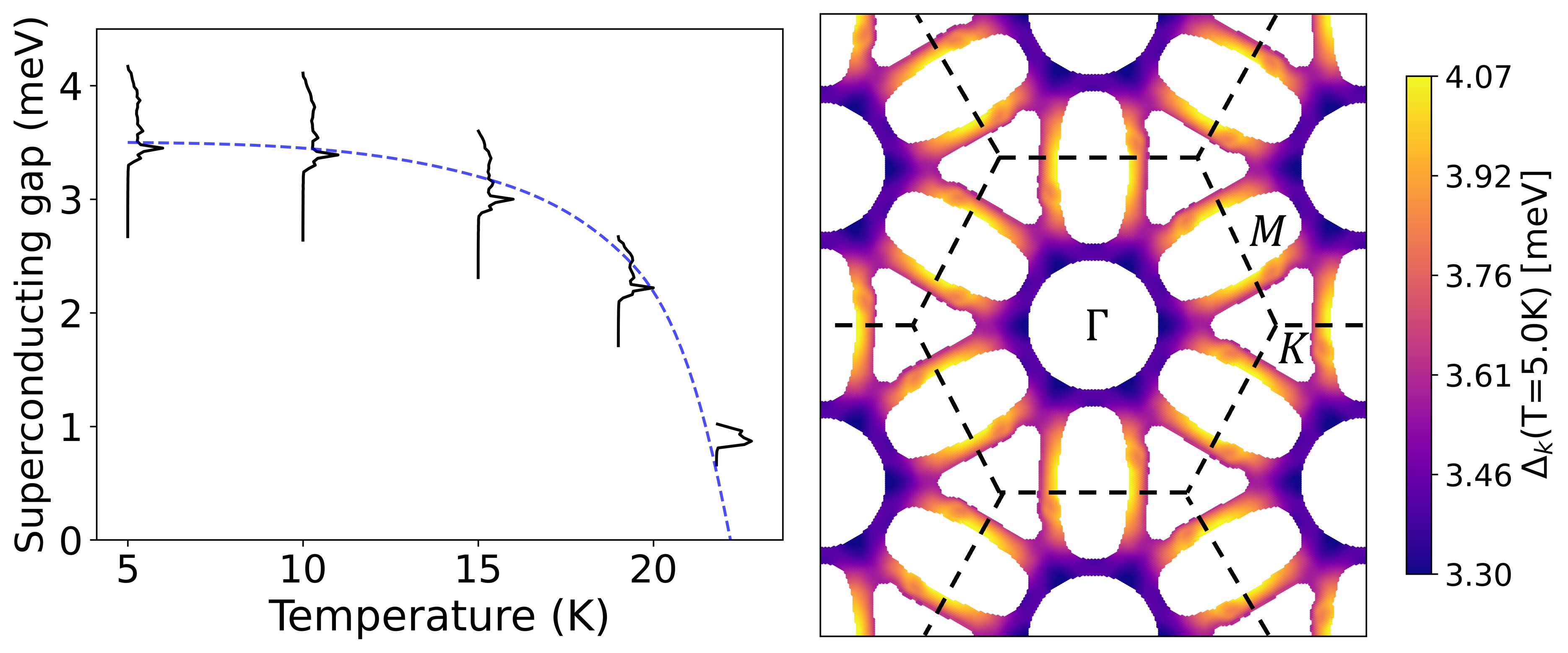}
		\caption{ Left: Temperature dependence of the superconducting gap $\Delta$ for Ti$_2$CSH, computed by solving the anisotropic Migdal-Eliashberg equations. The gap progressively decreases with increasing temperature and vanishes at 22.6~K, indicating the superconducting critical temperature $T_c$. Right: Momentum-resolved superconducting gap distribution $\Delta_k$ on the Fermi surface at 10~K. The color scale represents the magnitude of the gap, which ranges from 4.29 to 4.71~meV, illustrating the uniform and single-gap nature of the superconducting state.}
		\label{fig:epw-gap}
	\end{figure}

    Superconductivity mediated by electron-phonon interactions can be quantified through the electron-phonon coupling strength $\lambda_{n\boldsymbol{q}}$, as depicted in the EPC-weighted phonon dispersion in Figure~\ref{fig:phonons}. The most substantial contribution to the total coupling originates from low-energy phonon modes in the 0--50~meV range, which are primarily associated with vibrations of Ti and S atoms, contributing approximately 79\% to the overall coupling. In contrast, the intermediate energy modes between 60--75~meV, largely involving C atom vibrations, account for about 20\%. These contributions result in a total electron-phonon coupling constant of $\lambda = 0.79$. Other relevant superconducting quantities can also be computed; for example, the logarithmic average phonon energy is $\omega_{\text{log}} = 29.26$~meV, and the mean-square frequency is $\omega_2 = 40.43$~meV. 
    
    The superconducting transition temperature is identified by tracking the closure of the superconducting gap, which is obtained by solving the two coupled anisotropic Migdal-Eliashberg equations in their nonlinear form. The evolution of the gap with temperature, depicted in the left panel of Figure~\ref{fig:epw-gap}, shows single-gap superconducting behavior. Additionally, at 10.0~K, the gap distribution across the Fermi surface, illustrated in the right panel of Figure~\ref{fig:epw-gap}, reveals a uniform gap structure. The magnitude of the superconducting gap spans a range from approximately 4.29 to 4.71~meV, clearly indicating the presence of a single superconducting channel. In relation to the electronic structure, the superconducting gap of Ti$_2$CSH is primarily associated with Ti-$d$ orbital electrons, which undergo a superconducting phase transition to form Cooper pairs at the Fermi level. Therefore, the gap distribution mirrors the Fermi surface of the electrons, with the contours, illustrated in the right panel of Figure~\ref{fig:epw-gap}, indicating the strength of the binding energy of phonon-mediated Cooper pairs. As the temperature rises, the superconducting gaps $\Delta(K)$ gradually decrease and ultimately vanish at 22.6~K, which defines the superconducting critical temperature. 
    
    Hydrogenation has emerged as a powerful route to tune the electronic and magnetic properties of 2D transition metal dichalcogenides (TMDs). Several Janus hydrides, including MoSH, WSH, and WSeH, have been predicted or observed to exhibit superconductivity with $T_c$ values in the 12–30~K range~\cite{lu2017janus,liu2022two,ku2023ab,wseh_prb,wsh_2dmat,gan2024hydrogenation,fu2024superconductivity}. More recently, theoretical studies have extended these predictions to Ti- and Zr-based Janus chalcogenide hydrides, where superconductivity is likewise reported in the same temperature window~\cite{li2024machine,ul2024superconductivity}. Within this broader context, our calculations reveal that Ti$_2$CSH is a phonon-mediated superconductor with an electron–phonon coupling constant $\lambda = 0.79$ and a critical temperature $T_c = 22.6$~K. The single, uniform superconducting gap, dominated by Ti-$d$ orbital states. Overall, the predicted $T_c$ of Ti$_2$CSH is competitive with other Janus hydrides, placing it firmly within the growing family of hydrogenated Janus superconductors and highlighting its potential as a stable 2D platform for multifunctional applications.  
    
	\section{Conclusion}
In summary, we have conducted a comprehensive first-principles investigation of the structural, vibrational, and superconducting properties of the Ti$_2$CSH monolayer. Phonon dispersion calculations confirm its dynamic stability, with all phonon frequencies being real and positive across the Brillouin zone. The detailed analysis of phonon modes at the $\Gamma$ point reveals the system belongs to the $C_{3v}$ point group, with all optical modes being both Raman and infrared active due to the lack of inversion symmetry. The phonon density of states highlights the atomic contributions across energy ranges: low-energy vibrations are dominated by Ti and S atoms, mid-range by C atoms, and high-frequency modes by H atoms.

Notably, significant electron-phonon coupling (EPC) is observed, especially in the low-energy acoustic and optical branches, leading to a total EPC constant of $\lambda = 0.79$. This strong EPC supports phonon-mediated superconductivity. Solving the anisotropic Migdal-Eliashberg equations reveals a single-gap superconducting state with a superconducting gap magnitude ranging from 4.29 to 4.71~meV at 10~K. The temperature dependence of the gap indicates a superconducting critical temperature $T_c$ of approximately 22.6~K, above the boiling point of liquid hydrogen.

Our findings suggest that Ti$_2$CSH is a dynamically stable, phonon-mediated superconductor with a relatively high $T_c$ among 2D materials. This makes it a promising candidate for low-dimensional superconducting devices and quantum technologies. Further experimental synthesis and validation are encouraged to explore its practical applications.


    \section*{Data Availability}
    The data that support the findings of this study are available from the corresponding
    authors upon reasonable request.
    
    \section*{Code Availability}
    The first-principles DFT calculations were performed using the open-source Quantum ESPRESSO package, available at \url{https://www.quantum-espresso.org}, along with pseudopotentials from the Quantum ESPRESSO pseudopotential library at \url{https://pseudopotentials.quantum-espresso.org/}. Electron-phonon coupling and related properties were computed using the EPW code, available at \url{https://epw-code.org/}.

    \section*{Acknowledgements}
	This research project is supported by the Second Century Fund (C2F), Chulalongkorn University. We acknowledge the supporting computing infrastructure provided by NSTDA, CU, CUAASC, NSRF via PMUB [B05F650021, B37G660013] (Thailand). URL:www.e-science.in.th.

    \section*{Author Contributions}
    Jakkapat Seeyangnok performed all of the calculations, analysed the results, wrote the first draft manuscript, and coordinated the project. Udomsilp Pinsook analysed the results and wrote the manuscript.

    \section*{Conflict  of Interests}
    The authors declare no competing financial or non-financial interests.

    \section*{Appendix A: Wannier interpolation}
    \begin{figure}[h!]
		\centering  
    	\includegraphics[width=8cm]{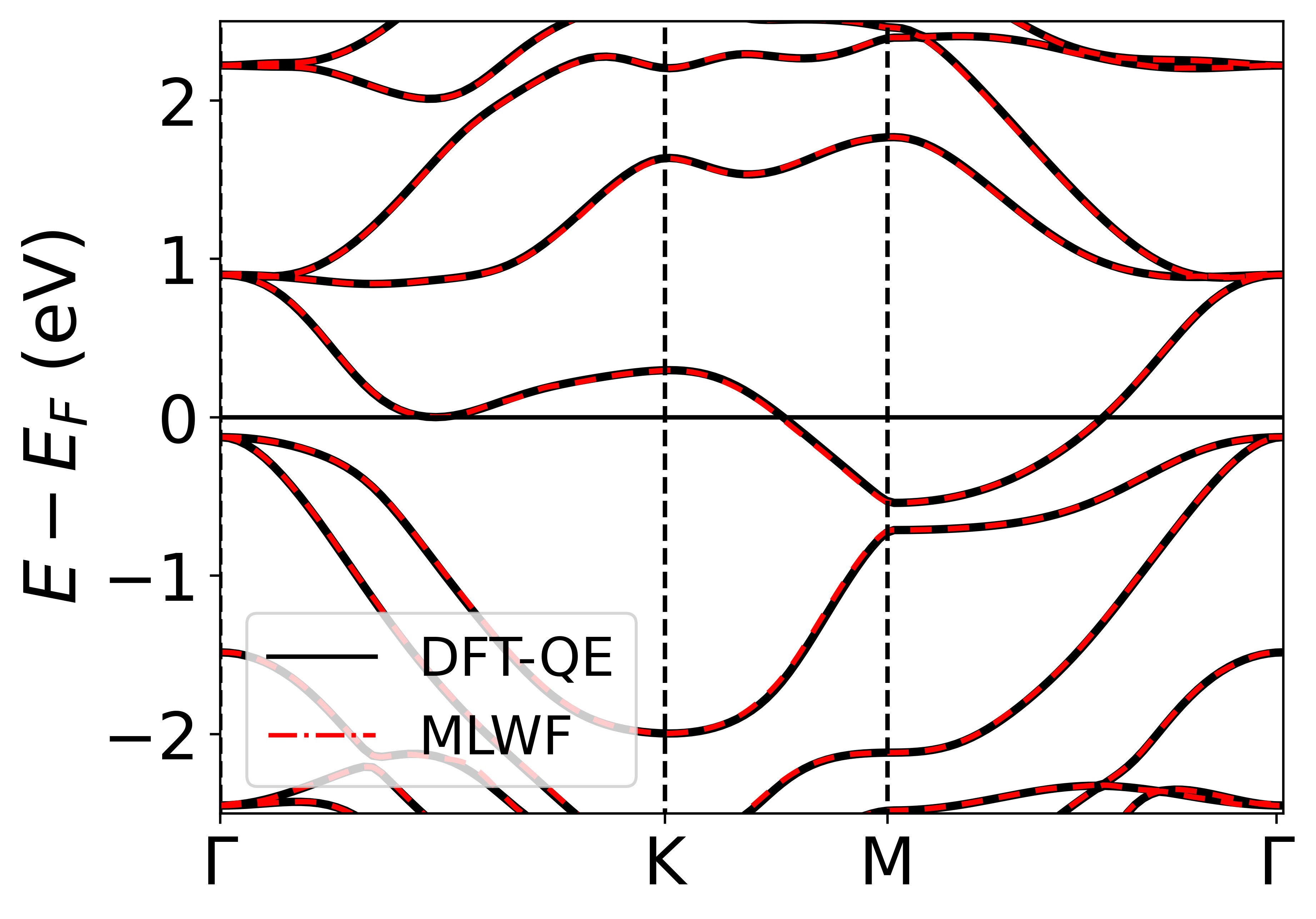}
		\caption{Figure shows a side-by-side comparison of electronic band structures derived from conventional DFT methods (in black) and those reconstructed using maximally localized Wannier functions (dashed red). They also depict the Fermi surface near the Fermi energy, a critical feature for evaluating electron-phonon coupling in the context of anisotropic Migdal-Eliashberg theory.}
		\label{fig:epw}
    \end{figure}  
    Figure~\ref{fig:epw} displays the maximally localized Wannier functions (MLWFs), obtained using the EPW software. These functions provide a compact and efficient representation of the electronic structure, enabling high-resolution interpolation of both the band energies and electron-phonon coupling elements throughout the Brillouin zone. Their strong spatial localization is key to accurately capturing the electron-phonon interaction, which underpins the calculation of the Eliashberg spectral function and the resulting superconducting characteristics.
    
    \bibliographystyle{unsrt} 
	\bibliography{references}

\end{document}